\documentclass[fleqn]{mat01}
\usepackage{times,mathtimy,amssymb,latexsym,epsfig}
\begin{document}

\makeatletter
\def\artpath#1{\def\@artpath{#1}}
\makeatother \artpath{d:/prema/pm-crc}

\newtheorem{coro}[defin]{\rm COROLLARY}
\newtheorem{exam}[defin]{\it Example}

\newtheorem{theore}{Theorem}
\renewcommand\thetheore{\arabic{section}.\arabic{theore}}
\newtheorem{pot}[theore]{\it Proof of Theorem}

\def\d{\mbox{\rm d}}
\def\e{\mbox{\rm e}}
\def\Re{\mbox{\rm Re}}
\def\sign{\mbox{\rm sign}}
\def\Ran{\mbox{\rm Ran}}

\setcounter{page}{375} \firstpage{375}

\renewcommand\theequation{\arabic{section}.\arabic{equation}}

\newcommand{\aref}[1]{Assumption~{{\rm\ref{#1}}}}
\newcommand{\tref}[1]{Theorem~{\ref{#1}}}
\newcommand{\cref}[1]{Corollary~{\ref{#1}}}
\newcommand{\lref}[1]{Lemma~{\ref{#1}}}
\newcommand{\rref}[1]{Remark~{\ref{#1}}}
\newcommand{\eref}[1]{Example~{\ref{#1}}}
\newcommand{\pref}[1]{Proposition~{\ref{#1}}}
\newcommand{\bx}{\mathbf{x}}
\newcommand{\by}{\mathbf{y}}
\newcommand{\cH}{\mathcal{H}}
\newcommand{\cR}{\mathcal{R}}
\newcommand{\cO}{\mathcal{O}}
\newcommand{\cB}{\mathcal{B}}
\newcommand{\cN}{\mathcal{N}}
\newcommand{\cM}{\mathcal{M}}
\newcommand{\cK}{\mathcal{K}}
\newcommand{\bC}{\mathbf{C}}
\newcommand{\bR}{\mathbf{R}}
\newcommand{\bN}{\mathbf{N}}
\newcommand{\jap}[1]{\langle{#1}\rangle}
\newcommand{\ip}[2]{\langle{#1},{#2}\rangle}
\newcommand{\ket}[1]{\lvert{#1}\rangle}
\newcommand{\bra}[1]{\langle{#1}\rvert}
\newcommand{\ka}{\kappa}
\newcommand{\Tt}{\widetilde{T}}
\newcommand{\ft}{\widehat{\phantom{\cdot}}}

\newcommand{\dint}{\int\!\!\!\!\int_{\mathbf{R}^{2}}}
\newcommand{\rr}{\mathbf{R}}

\newcommand{\hl}{[0,\infty)}
\newcommand{\norm}[1]{\lVert{#1}\rVert}
\newcommand{\cD}{\mathcal{D}}
\newcommand{\bo}[2]{\mathcal{B}({#1},{#2})}
\newcommand{\wltwo}[1]{L^{2,{#1}}([0,\infty))}
\newcommand{\ltwo}{L^2([0,\infty))}
\newcommand{\bh}{\mathcal{B}(\cH)}
\newcommand{\vp}{\varphi}
\newcommand{\ve}{\varepsilon}
\newcommand{\goo}{G_0^D}


\markboth{Arne Jensen and Gheorghe Nenciu}{Schr\"{o}dinger operators on the half line}

\title{Schr\"{o}dinger operators on the half line: Resolvent expansions and the
Fermi golden rule at thresholds}

\author{ARNE JENSEN$^{*}$ and~GHEORGHE NENCIU$^{\dagger, \ddagger}$}

\address{$^{*}$Department of Mathematical Sciences, Aalborg University, Fredrik Bajers Vej 7G,
DK-9220 Aalborg \O{}, Denmark\\
\noindent $^{\dagger}$Department of Theoretical Physics, University of Bucharest, P.
O. Box
MG11, 76900~Bucharest, Romania\\
\noindent $^{\ddagger}$Institute of Mathematics ``Simion Stoilow'' of the Romanian
Academy, P. O. Box 1-764, RO-014700 Bu\-cha\-rest, Romania\\
\noindent E-mail: matarne@math.aau.dk; nenciu@barutu.fizica.unibuc.ro;
Gheorghe.Nenciu@imar.ro\\[1.2pc]
\noindent {\it Dedicated to K B Sinha on the occasion of his
sixtieth \vspace{-1pc}birthday}}

\volume{116}

\mon{November}

\parts{4}

\pubyear{2006}

\Date{}

\begin{abstract}
We consider Schr\"{o}dinger operators $H=- \d^2/\d r^2+V$ on $L^2([0,\infty))$ with
the Dirichlet boundary condition. The potential $V$ may be local or non-local, with
polynomial decay at infinity. The point zero in the spectrum of $H$ is classified, and
asymptotic expansions of the resolvent around zero are obtained, with explicit
expressions for the leading coefficients. These results are applied to the
perturbation of an eigenvalue embedded at zero, and the corresponding modified form of
the Fermi golden\break rule.
\end{abstract}

\keyword{Schr\"{o}dinger operator; threshold eigenvalue; resonance; Fermi golden
rule.}

\maketitle

\section{Introduction}\label{sect1}

This paper is a continuation of \cite{JenNen,JN-FGR}, where
expansions of the resolvents of Schr\"odinger type operators at
thresholds, as well as the form of the Fermi golden rule (which
actually goes back to Dirac), when perturbing a nondegenerate
threshold eigenvalue, were obtained. While the methods and results
in \cite{JenNen,JN-FGR} are to a large extent abstract, the
examples discussed were restricted to Schr\"{o}dinger operators in
odd dimensions with local potentials. The aim of this paper is to
show that the methods in \cite{JenNen,JN-FGR} allow to treat the
non-local potentials in exactly the same manner as the local ones,
although the properties of the corresponding operators can be
quite different. For example, one can have zero as an eigenvalue
in one dimension, or eigenfunctions for the zero eigenvalue with
compact support (in this connection see e.g.~\cite{amrein}).

Let us briefly describe the results. Let $H_0^D$ denote $-{\d^2}/{\d r^2}$ on
$\cH=L^2([0,\infty))$ with the Dirichlet boundary condition. Let $V$ be a potential,
which can be either local or non-local. We assume that $V$ is a bounded selfadjoint
operator on $\cH$. Let $\cH^s=L^{2,s}([0,\infty))$ denote the weighted space. Then we
assume that $V$ extends to a bounded operator from $\cH^{-\beta/2}$ to $\cH^{\beta/2}$
for a sufficiently large $\beta>0$. Since we are concerned with threshold phenomena,
the first step is to study the solutions of the equation $H\Psi=0$. The result is that
under the above conditions, for the solutions of $H\Psi=0$ there are four\break
possibilities:

\begin{enumerate}
\renewcommand\labelenumi{(\roman{enumi})}
\leftskip .3pc
\item No non-zero solutions. In this case zero is called a \emph{regular} point for $H$.

\item One non-zero solution in $L^{\infty}([0,\infty))$, but not in $L^2([0,\infty))$.
In this case zero is called an \emph{exceptional} point of the \emph{first} kind for
$H$.

\item A finite number of linearly independent solutions, all belonging to
$L^2([0,\infty))$. In this case zero is called an \emph{exceptional} point of the
\emph{second} kind for $H$.

\item Two or more linearly independent solutions, which can be chosen such that all but one
belong to $L^2([0,\infty))$. In this case zero is called an \emph{exceptional} point
of the \emph{third} kind for $H$.
\end{enumerate}
Let us note that if $V$ is multiplication by a function, i.e.
$(Vf)(r) = \mathsf{V}(r)f(r)$ for some function $\mathsf{V}(r)$,
then only cases (i) and (ii) occur.

In all cases we obtain asymptotic expansions for the resolvent of $H$ around
the point zero. It is convenient to use the variable $\kappa=-i\sqrt{z}$ in
these expansions. We have
\begin{equation*}
(H+\kappa^2)^{-1}=\sum_{j=-2}^p\kappa^jG_j+\cO(\kappa^{p+1})
\end{equation*}
as $\kappa\to0$, in the topology of the bounded operators from $\cH^s$ to
$\cH^{-s}$ for a sufficiently large $s$, depending on $p$ and the classification
of the point zero for $H$. We compute a few of the leading coefficients
explicitly.

These results on asymptotic expansion for the resolvent, and the explicit expressions
for the coefficients, are the main ingredients for the application of the results in
\cite{JN-FGR}, concerning the perturbation of an eigenvalue embedded at the threshold
zero. The main result from \cite{JN-FGR} in the context of the Schr\"{o}dinger
operators on the half line considered above is as follows. Let $H=H_0^D+V$, where $V$
satisfies Assumption~\ref{nonlocalV} for a sufficiently large $\beta$. Let $W$ be
another potential satisfying the same assumption. We consider the family $H(\ve)=H+\ve
W$ for $\ve>0$. Assume that $0$ is a simple eigenvalue of $H$, with normalized
eigenfunction $\Psi_0$. Assume
\begin{equation}\label{def-b2}
b=\ip{\Psi_0}{W\Psi_0}>0,
\end{equation}
and that for some odd integer $\nu\geq-1$ we have
\begin{equation}\label{G-cond2}
G_{j}=0,\quad \text{for $j=-1,1,\ldots,\nu-2$ \quad and} \quad
g_{\nu}=\ip{\Psi_0}{WG_{\nu}W\Psi_0}\neq0.
\end{equation}

Then Theorem~3.7 in  \cite{JN-FGR} gives the following result (the
modified Fermi golden rule) on the survival probability for the
state $\Psi_0$ under the evolution $\exp(-itH(\ve))$, showing that
for $\ve$ sufficiently small the eigenvalue zero of $H$ becomes a
resonance.

There exists $\ve_0>0$, such that for $0<\ve<\ve_0$ we have
\begin{equation}
\ip{\Psi_0}{\e^{-itH(\ve)}\Psi_0}=\e^{-it\lambda(\ve)}+\delta(\ve,t),\quad t>0.
\end{equation}
Here $\lambda(\ve)=x_0(\ve)-i\Gamma(\ve)$ with
\begin{align}
\Gamma(\ve)&=-i^{\nu -1}g_{\nu}b^{\nu/2}\ve^{2+(\nu/2)}(1+\mathcal{O}(\ve)),\\[.3pc]
x_{0}(\ve)&=b\ve(1+\mathcal{O}(\ve)),
\end{align}
as $\ve\to0$. The error term satisfies
\begin{equation}
|\delta(\ve,t)|\leq C \ve^{p(\nu)},\quad t>0, \quad p(\nu)=\min\{2,(2+\nu)/2\}.
\end{equation}
As an application of the results on asymptotic expansion of the resolvent of $H$ near
zero we explicitly compute the coefficient $g_{\nu}$ in two cases.

The contents of the paper is as follows: In \S2 we introduce some
notation used in the rest of the paper. Section~3 forms the core
of the paper and contains our results on the resolvent expansions
for the free Schr\"odinger operator on the half line, and then for
the Schr\"odinger operator with a general class of potentials,
including non-local ones. In \S4 we illustrate the general results
by giving an explicit example with a rank 2 operator as the
perturbation. Finally, \S5 contains the results on the modified
Fermi golden rule for the class of operators considered here.

Let us conclude with some remarks on the literature. Resolvent expansions of the type
obtained here are typical for Schr\"{o}dinger operators in odd dimensions, when the
potential decays rapidly. Such results were obtained in
\cite{jensen-kato79,jensen80,murata82}. More recently, a unified approach was
developed in \cite{JenNen,JenNen-err}. It is this approach that we use here. Another
approach to the threshold behavior is to use the Jost function. See for example
\cite{aktosun00,yafaev82}. See also the cited papers for further references to results
on resolvent expansions around thresholds.

\section{Notation}
\setcounter{equation}{0}

Let $H$ be a self-adjoint operator on a Hilbert space $\cH$. Its
resolvent is denoted by $R(z)=(H-z)^{-1}$. In the sequel we will
often look at operators with essential spectrum equal to
$[0,\infty)$, such that $0$ is a threshold point. We will look at
asymptotic expansions around this point for the resolvent. It is
convenient to change the variable $z$ by introducing
$z=-\kappa^2$, with $\Re\,\kappa>0$.

In the half line case there is a type of notation common in the physics
literature that is very convenient. The resolvent will have an integral kernel
$k(r,r')$, $r,r'\in\hl$. We introduce the two functions
\begin{equation}\label{r-rprime}
r_>=\max\{r,r'\},\quad r_<=\min\{r,r'\}.
\end{equation}
We note a few properties for future reference
\begin{equation}
r_> + r_< = r + r', \quad r_> - r_< = |r-r'|,\quad r_>\cdot r_<=r\cdot r'.
\end{equation}

The weighted $L^2$-space on the half line is given by
\begin{equation}
\cH^s=\wltwo{s}=\left\{f\in L^2_{\rm loc}([0,\infty)) \,\bigg|\,\int_0^{\infty}
|f(r)|^2(1+r^2)^s \d r<\infty\right\},
\end{equation}
for $s\in\rr$. We write $\cH=\cH^0=L^2([0,\infty))$.
We use the notation $\bo{s_1}{s_2}$ for the bounded operators from $\cH^{s_1}$
to $\cH^{s_2}$.

The inner product $\ip{\cdot}{\cdot}$ on $\cH$ is also used to denote the duality
between $\cH^s$ and $\cH^{-s}$. We use the bra and ket notation for operators from
$\cH^{s}$ to $\cH^{-s}$. For example, the operator $f\mapsto \int_0^{\infty}f(r) \d
r\cdot 1$ from $\cH^{s}$ to $\cH^{-s}$ for $s>1/2$ is denoted by $\ket{1}\bra{1}$.

In the asymptotic expansions below there will be error terms in
the norm topology of $\bo{s_1}{s_2}$ for specified values of the
parameters $s_1$ and $s_2$. Here $\kappa\in\{\zeta\,|\,0<
|\zeta|<\delta,\,\Re\,{\zeta}>0\}$ for a sufficiently small
$\delta$. We will use the standard notation $\cO(\kappa^p)$ for
these error terms.

\section{Resolvent expansions}
\setcounter{equation}{0}

In this section we first obtain the resolvent expansion of the free Schr\"{o}dinger
operator on the half line, and then for the Schr\"{o}dinger operator with a general
class of potentials, including non-local ones.

\subsection{\it The free operator with the Dirichlet boundary condition}\label{dirichlet}

We denote by $H_0^D$ the operator with the domain and action given by
\begin{equation}\label{DBC-domain}
\cD(H_0^D)=\{f\in\cH\,|\,f\in AC^2([0,\infty)),\: f(0)=0\},\quad
H_0^Df=-\frac{\d^2}{\d r^2}f.
\end{equation}
Here the space $AC^2$ denotes functions $f$ that are continuously differentiable on
$[0,\infty)$, with $f'$ absolutely continuous (see \cite{rs1}). It is well-known that
this operator is self-adjoint.

The resolvent $R_0^D(z)=(H_0^D-z)^{-1}$ has the integral kernel (using $z=-\kappa^2$
as above)
\begin{equation}
K_0^D(\kappa;r,r')=-\frac{i}{\kappa}\sin(i\kappa r_<)\e^{-\kappa r_>},
\end{equation}
which can be rewritten as
\begin{equation}
K_0^D(\kappa;r,r')=-\frac{1}{2\kappa} (\e^{-\kappa(r_>+r_<)}-\e^{-\kappa(r_>-r_<)}).
\end{equation}
Using the Taylor expansion we can get the following result, as in
\cite{jensen-kato79,jensen80,murata82}.

\begin{Proposition}\label{prop1}$\left.\right.$\vspace{.5pc}

\noindent The resolvent $R_0^D(-\kappa^2)$ has the following asymptotic expansion. Let
$p\geq0$ be an integer and let $s>p+\frac32$. Then we have
\begin{equation}\label{Dirichlet-asymp}
R_0^D(-\kappa^2)=\sum_{j=0}^{p}G^D_j\kappa^j+\cO(\kappa^{p+1})
\end{equation}
in the norm topology of $\bo{s}{-s}$. The operators $G^D_j$ are given explicitly in
terms of their integral kernels by
\begin{equation}\label{g00}
G^D_j\!: \frac{(-1)^j}{2(j+1)!}((r_>+r_<)^{j+1}-(r_>-r_<)^{j+1}).
\end{equation}
Let $s_1,s_2>\frac12$ with $s_1+s_2>2$. Then $G_0^D\in\bo{s_1}{-s_2}$. For
$s>\frac32$ we also have $G^D_0\in\cB(\cH^s,L^{\infty}([0,\infty)))$.

If $j\geq1$ and  $s>j+\frac12${\rm ,} then $G^D_j\in\bo{s}{-s}$.
\end{Proposition}

\begin{proof}
The straightforward computations and estimates are omitted.
\end{proof}

\begin{remark}{\rm For future reference we note the expressions
\begin{align}
&G^D_0\!: r_<,\label{dg0}\\[.3pc]
&G^D_1\!: -r_<r_>=-r\cdot r',\label{dg1}\\[.3pc]
&G^D_2\!: \frac{1}{2}r_<r_>^2+\frac{1}{6}r_<^3\label{dg2},\\[.3pc]
&G^D_3\!:-\frac{1}{6}(r_>^3r_<+r_>r_<^3)=-\frac{1}{6}(r^3\cdot r'
+r\cdot(r')^3).\label{dg3}
\end{align}}
\end{remark}

\subsection{\it The potential and the factorization method}

We now add a potential $V$ to $H_0^D$ and find the asymptotic expansion of the
resolvent of $H=H_0^D+V$ around zero. We will allow a rather general class of
potentials, so we introduce the following assumption. We consider only bounded
perturbations, however it is possible to extend the results to potentials with
singularities.

\setcounter{defin}{2}
\begin{assumption}\label{nonlocalV}
{\rm Let $V$ be a bounded self-adjoint operator on $\cH$, such that $V$ extends to a
bounded operator from $\cH^{-\beta/2}$ to $\cH^{\beta/2}$ for some $\beta>2$. Assume
that there exists a Hilbert space $\cK$, a compact operator
$v\in\cB(\cH^{-\beta/2},\cK)$, and a self-adjoint operator $U\in\cB(\cK)$ with
$U^2=I$, such that $V=v^*Uv$.}
\end{assumption}

\begin{remark}\label{remark3}
{\rm The factorization leads to a natural additive structure on the potentials. Assume
that $V_j=v_j^*U_jv_j$, $j=1,2$, satisfy \aref{nonlocalV}. Let $\cK=\cK_1\oplus\cK_2$.
Using matrix notation we define
\begin{equation}\label{sumfactor}
v=\begin{bmatrix}
v_1\\
v_2
\end{bmatrix},
\qquad
U=\begin{bmatrix}
U_1 & 0 \\
0 & U_2
\end{bmatrix}.
\end{equation}
Then it follows that $V=V_1+V_2$ has the factorization $V=v^*Uv$ with the operators
$v$ and $U$ defined in \eqref{sumfactor} and the space $\cK=\cK_1\oplus\cK_2$.}
\end{remark}

\begin{exam}\label{ex34}
{\rm We give two examples, the first one a local perturbation, and the second one a
non-local perturbation.

\begin{enumerate}
\renewcommand\labelenumi{\rm (\roman{enumi})}
\leftskip .3pc

\item Let $V$ be multiplication by a real-valued function $\mathsf{V}(r)$. Assume that
\begin{equation*}
|\mathsf{V}(r)|\leq C(1+r)^{-\beta}
\end{equation*}
for some $\beta>2$. Take $\cK=\cH$ and let $v=v^*$ denote multiplication by
$|\mathsf{V}(r)|^{1/2}$. Let $U$ denote multiplication by $1$, if
$\mathsf{V}(r)\geq0$, and by $-1$, if $\mathsf{V}(r)<0$. Then all conditions in
\aref{nonlocalV} are satisfied.

\item Let $\vp\in\cH^{\beta/2}$ and $\gamma\in\rr$, $\gamma\neq0$. Let
$V=\gamma\ket{\vp}\bra{\vp}$. It has the following factorization. Let $\cK=\bC$. Let
$v\colon\cH^{-\beta/2}\to\cK$ be given by $v(f)=|\gamma|^{1/2}\ip{\vp}{f}$, and $U$
multiplication by $\sign(\gamma)$. Then $v^*(z)=z|\gamma|^{1/2}\vp$, and we have
$V=v^*Uv$. The generalization to an operator of rank $N$ follows from
Remark~\ref{remark3}.
\end{enumerate}}
\end{exam}

Write $H=H_0^D+V$ with $V$ satisfying \aref{nonlocalV}.
We note the following result.

\begin{lemma}
Let $V$ satisfy \aref{nonlocalV}. Then $V$ is $H_0^D$-compact.
\end{lemma}

\begin{proof}
We have
\begin{equation*}
V(H_0^D+i)^{-1}=[V(1+r)^{\beta/2}][(1+r)^{-\beta/2}(H_0^D+i)^{-1}].
\end{equation*}
The first factor $[\cdots]$ is bounded by the assumption and the
second factor $[\cdots]$ is compact by well-known arguments.
\hfill $\Box$
\end{proof}

We now briefly recall the factorization method, as used in \cite{JenNen}, but
here extended to cover the non-local potentials.
The starting point is the operator
\begin{equation*}
M(\kappa)=U+v(H_0^D+\kappa^2)^{-1}v^*,
\end{equation*}
which is now a bounded operator on $\cK$. The factored second resolvent
equation is given by
\begin{equation}\label{freq}
R(-\kappa^2)=R_0^D(-\kappa^2)-R_0^D(-\kappa^2)
v^*M(\kappa)^{-1}vR_0^D(-\kappa^2).
\end{equation}

The first step in obtaining an asymptotic expansion for $ R(-\kappa^2)$ is to
study the invertibility of $M(\kappa)$ and the asymptotic expansion of the
inverse. Inserting the asymptotic expansion
\eqref{Dirichlet-asymp} we get
\begin{equation}\label{M-asymp}
M(\ka)=\sum_{j=0}^p\ka^jM_j+\cO(\ka^{p+1}),
\end{equation}
provided $\beta>2p+3$. Here
\begin{equation}\label{M-coeff}
M_0=U+vG_0^Dv^*\quad\text{and}\quad M_j=vG_j^Dv^*,\quad j=1,\ldots,p.
\end{equation}

\subsection{\it Analysis of $\ker M_0$}

We analyze the structure of $\ker M_0$ and the connection with the point
zero in the spectrum of $H$.

\setcounter{defin}{6}
\begin{lemma}\label{NL-lemma}
Let \aref{nonlocalV} be satisfied with $\beta>3$.

\begin{enumerate}
\renewcommand\labelenumi{\rm (\roman{enumi})}
\leftskip .3pc

\item Let $f\in\ker M_0$. Define $g=-G^D_0v^*f$. Then $Hg=0,$ with the derivatives
in the sense of distributions. We have that $g\in L^{\infty}([0,\infty))\cap C(\hl),$
with $g(0)=0$. We have $g\in\cH,$ if and only if
\begin{equation}\label{nonlocalltwo}
\ip{vr}{f}_{\cK}=0.
\end{equation}

\item Assume $g\in\cH^{-s}\cap C(\hl),$ $s\leq3/2,$ satisfies $g(0)=0$ and
$Hg=0,$ in the sense of distributions. Let $f=Uvg$. Then $f\in\ker M_0$.

\item Assume additionally that $V$ is multiplication by a function.
Let $f\in\ker M_0,$ $f\neq0$. Then $\ip{vr}{f}\neq0,$ and $\dim\ker M_0=1$.
\end{enumerate}
\end{lemma}

\begin{proof}
Let $f\in\ker M_0$, and define $g=-G^D_0v^*f$. Then we have
\begin{equation*}
g(r)=-\int_0^{\infty}r'(v^*f)(r')\d r'-\int_r^{\infty}(r-r')(v^*f)(r')\d r'.
\end{equation*}
Since $v^*f\in\cH^{s}$ for some $s>3/2$, the second term belongs to $\cH$. The first
term is a constant. Thus part (i) follows. For part (ii), assume $g\in\cH^{-s}\cap
C(\hl)$, $s\leq3/2$, satisfies $g(0)=0$ and $Hg=0$, in the sense of distributions.
Then $f=Uvg\in\cK$. By assumption and definition we have
\begin{equation*}
\frac{\d^2}{\d x^2}g=Vg=v^*f.
\end{equation*}
The mapping properties of $v^*$ imply that $v^*f\in\cH^s$ for some $s>3/2$.
Thus we can define
\begin{equation*}
h(r)=-\int_r^{\infty}(r-r')(v^*f)(r')\d r'.
\end{equation*}
Hence
\begin{equation*}
\frac{\d^2}{\d x^2}h=v^*f.
\end{equation*}
We conclude that $\frac{{\rm d}^2}{{\rm d} r^2}(h-g)=0$ in the
sense of distributions, and thus for some $a,b\in\bC$ we have
$g(r)=h(r)+a+br$. Since $g\in\cH^{-s}$, $s\leq3/2$, and $h\in\cH$,
we conclude that $b=0$. Since $g(0)=0$ by assumption, we have
\begin{equation*}
a=-h(0)=-\int_0^{\infty}r'v(r')f(r')\d r.
\end{equation*}
Thus we have shown that
\begin{equation*}
g(r)=-\int_0^{\infty}r'v(r')f(r')\d r-\int_r^{\infty}(r-r')(v^*f)(r')\d r'=
-(G_0^Dv^*f)(r),
\end{equation*}
such that
\begin{equation*}
Uf=UUvg=vg=-vG_0^Dv^*f,
\end{equation*}
or $M_0f=0$.

Assume now that $V$ is multiplication by a function $\mathsf{V}$, and that the
factorization is chosen as above in Example~\ref{ex34}. To prove part (iii), assume
that $f\in\ker M_0$ and that $\ip{vr}{f}=0$. Let $g=-G_0^Dv^*f$. Then $M_0f=0$ implies
$f=Uvg$. Using $\ip{vr}{f}=0$, we find that $g$ satisfies the homogeneous Volterra
equation
\begin{equation*}
g(r)=-\int_r^{\infty}(r-r')\mathsf{V}(r')g(r')\d r'.
\end{equation*}
It follows by a standard iteration argument that $g=0$,  and then also $f=0$. To prove
the final statement, assume that we have $f_j\in\ker M_0$, and $f_j\neq0$, $j=1,2$.
Define $g_j=-G_0^Dv^*f_j$. Then we can find $\alpha\in\bC$, such that
$\ip{vr}{f_1}+\alpha\ip{vr}{f_2}=0$. Thus we get
\begin{equation*}
(g_1+\alpha g_2)(r)=-\int_r^{\infty}(r-r')\mathsf{V}(r')(g_1+\alpha g_2)(r')\d r'.
\end{equation*}
It follows again by the iteration argument that $g_1+\alpha
g_2=0$, and then as above also $f_1+\alpha f_2=0$. This concludes
the proof of part (iii).
\end{proof}

\begin{remark}
{\rm Let us note that for a local potential it suffices to assume $\beta>2$ for the
results in Lemma~\ref{NL-lemma} to hold, since in this case we can use the mapping
property of $G_0^D$ given in Proposition~\ref{prop1}.}
\end{remark}

We need the following result, which is analogous to Lemma~2.6 of \cite{jensen-kato79}.
We include the proof here.

\begin{lemma}\label{g2lemma}
Assume that $f_j\in\cK$, such that \eqref{nonlocalltwo} holds for
$f_j$, $j=1,2$. Then we have that
\begin{equation}
\ip{f_1}{vG^D_2v^*f_2}=-\ip{G_0^Dv^*f_1}{G_0^Dv^*f_2}.
\end{equation}
\end{lemma}

\begin{proof}
Let $g_j=-G_0^Dv^*f_j$. Since \eqref{nonlocalltwo} holds, we have that
$g_j\in\ltwo$. Furthermore, we have
\begin{equation}\label{eq}
\frac{\d^2}{\d r^2}g_j=v^*f_j
\end{equation}
in the sense of distributions. We denote the Fourier transform on
the line by \ $\widehat{\cdot}$. From~\eqref{eq} it follows that
we have
\begin{equation*}
\xi^2\widehat{g_j}(\xi)=-(v^*f_j)^{\widehat{\phantom{\cdot}}}(\xi).
\end{equation*}
Since $v^*f_j\in\cH^s$ for some $s>3/2$, the Fourier transform
$(v^*f_j)^{\ft}$ is continuously differentiable, by the Sobolev embedding
theorem. Since $\widehat{g_j}\in L^2(\rr)$, we must have
\begin{equation}\label{zeroes}
(v^*f_j)^{\ft}(0)=0,\quad \frac{\d}{\d \xi}(v^*f_j)^{\ft}(0)=0.
\end{equation}
It follows from \eqref{dg1} that $G^D_1v^*f_j=0$. Thus we have \
\begin{equation*}
\ip{f_1}{vG^D_2v^*f_2}= \lim_{\kappa\to0}\frac{1}{\kappa^2}
\ip{v^*f_1}{((H_0^D+\kappa^2)^{-1}-G^D_0)v^*f_2}.
\end{equation*}
Now compute using the Fourier transform:
\begin{align*}
&\frac{1}{\kappa^2}\ip{v^*f_1}{((H_0^D+\kappa^2)^{-1}-G^D_0)v^*f_2}\\[.3pc]
&\quad\,=\frac{1}{\kappa^2}\int_{-\infty}^{\infty}\overline{(v^*f_1)^{\ft}(\xi)}
\left( \frac{1}{\xi^2+\kappa^2}-\frac{1}{\xi^2}
\right)(v^*f_2)^{\ft}(\xi)\d \xi.\\[.3pc]
&\quad\,=\int_{-\infty}^{\infty}\overline{(v^*f_1)^{\ft}(\xi)}
\frac{-1}{(\xi^2+\kappa^2)\xi^2} (v^*f_2)^{\ft}(\xi)\d \xi.
\end{align*}
It follows from \eqref{zeroes} that
\begin{equation*}
\frac{1}{\xi^2}(v^*f_j)^{\ft}(\xi)\in L^2(\rr).
\end{equation*}
Thus we can use dominated convergence and take the limit
$\kappa\to0$ under the integral sign above, to get the result.
\end{proof}

\subsection{\it Resolvent expansions{\rm :} Results}\label{results}

Let us now state the results obtained. We use the same terminology as
in~\cite{jensen-kato79}, since we have the same four possibilities for the point zero.
We say that zero is a \emph{regular point} for $H$, if $\dim\ker M_0=0$. We say that
zero is an \emph{exceptional point of the first kind}, if $\dim\ker M_0=1$, and there
is an $f\in\ker M_0$ with $\ip{vr}{f}\neq0$. We say that zero is an \emph{exceptional
point of the second kind}, if $\dim\ker M_0\geq1$, and all $f\in\ker M_0$ satisfy
$\ip{vr}{f}=0$. In this case zero is an eigenvalue for $H$ of multiplicity $\dim\ker
M_0$. Finally, we say that zero is an \emph{exceptional point of the third kind}, if
$\dim\ker M_0\geq2$, and there is an $f\in\ker M_0$ with $\ip{vr}{f}\neq0$.

We introduce the following notation. Let $S$ denote the orthogonal projection
onto $\ker M_0$. Then $M_0+S$ is invertible in $\cB(\cK)$. We write
\begin{equation}\label{def-J0}
J_0=(M_0+S)^{-1}.
\end{equation}

\setcounter{defin}{9}
\begin{theorem}[\!]\label{theorem1}
Assume that zero is a \emph{regular} point for $H$. Let $p\geq1$ be an integer.
Assume that $\beta>2p+3$ and $s>p+\frac32$. Then we have the expansion
\begin{equation}
R(-\kappa^2)=\sum_{j=0}^p\ka^jG_j+\cO(\kappa^{p+1})
\end{equation}
in the topology of $\bo{s}{-s}$. We have
\begin{align}
G_0&=(I+G_0^DV)^{-1}G_0^D,\\[.3pc]
G_1&=(I+G_0^DV)^{-1}G_1^D(I+VG_0^D)^{-1}.
\end{align}
The kernels of the operators $G_0^D$ and $G_1^D$ are given in
\eqref{dg0} and \eqref{dg1}{\rm ,} respectively.
\end{theorem}

\begin{theorem}[\!]\label{thm-ex1}
Let $p\geq0$ be an integer{\rm ,} and let $V$ satisfy
\aref{nonlocalV} for some $\beta>2p+7$. Assume that zero is an
exceptional point of the \emph{first kind} for $H$. Assume that
$s>p+\frac72$. Then we have an asymptotic expansion
\begin{equation}\label{expand-first}
R(-\ka^2)=\sum_{j=-1}^p\ka^jG_j+\cO(\ka^{p+1})
\end{equation}
in the topology of $\bo{s}{-s}$. We have
\begin{equation}
G_{-1}=\ket{\Psi_c}\bra{\Psi_c},
\end{equation}
where
\begin{equation*}
\Psi_c=\frac{\ip{f}{vr}}{|\ip{f}{vr}|^2}G_0^Dvf,
\end{equation*}
for $f\in\ker M_0${\rm ,} $\|f\|=1$.
\end{theorem}

\begin{theorem}[\!]\label{thm-ex2}
Let $p\geq1$ be an integer{\rm ,} and let $V$ satisfy
\aref{nonlocalV} for some $\beta>2p+11$. Assume that zero is an
exceptional point of the \emph{second kind} for $H$. Assume that
$s>p+\frac{11}{2}$. Then we have an asymptotic expansion
\begin{equation}\label{expand-second}
R(-\ka^2)=\sum_{j=-2}^p\ka^jG_j+\cO(\ka^{p+1})
\end{equation}
in the topology of $\bo{s}{-s}$. We have
\begin{align}
G_{-2}&=P_0,\label{ex2-G-2}\\[.3pc]
G_{-1}&=0,\\[.3pc]
G_0&=\goo-\goo v^*J_0v\goo -\goo v^* J_0 v G_2^{D}VP_0
-P_0VG_2^Dv^*J_0v\goo\notag\\[.3pc]
&\quad\, +P_0VG_4^DVP_0+P_0VG_2^D+G_2^DVP_0,\\[.3pc]
G_1&=G_1^D-G_1^Dv^*J_0v\goo-\goo v^*J_0vG_1^D
 + G_3^DVP_0+P_0VG_3^D\notag \\[.3pc]
&\quad\, +G_1^Dv^* J_0vG_2^DVP_0+P_0VG_2^Dv^*J_0vG_1^D.\label{ex2-G1}
\end{align}
Here $P_0$ denotes the projection onto the zero eigenspace of
$H${\rm ,} and the operator $J_0$ is defined by \eqref{def-J0}.
\end{theorem}

\begin{theorem}[\!]\label{thm-ex3}
Let $p\geq0$ be an integer{\rm ,} and let $V$ satisfy
\aref{nonlocalV} for some $\beta>2p+11$. Assume that zero is an
exceptional point of the \emph{third kind} for $H$. Assume that
$s>p+\frac{11}{2}$. Then we have an asymptotic expansion
\begin{equation}\label{expand-third}
R(-\ka^2)=\sum_{j=-2}^p\ka^jG_j+\cO(\ka^{p+1})
\end{equation}
in the topology of $\bo{s}{-s}$. We have
\begin{align}
G_{-2}&=P_0,\label{G-2}\\[.3pc]
G_{-1}&=\ket{\Psi_c}\bra{\Psi_c}\label{G-1}.
\end{align}
Here $P_0$ is the orthogonal projection onto the zero
eigenspace{\rm ,} and $\Psi_c$ is the canonical zero resonance
function defined in \eqref{resonancefcn}.
\end{theorem}

\begin{remark}
{\rm It is instructive to compare the results above with the results in the case of
dimension $d=3$ (see \cite{jensen-kato79}). The operator we consider here is the
angular moment component $\ell=0$ of $-\Delta+V$ on $L^2(\rr^3)$, provided $V$
commutes with rotations. In particular, we can only get zero as an eigenvalue for
non-local $V$, and the expansion in the second exceptional case has coefficient
$G_{-1}=0$ (and in the third exceptional case this coefficient only contains the zero
resonance term), consistent with the result in \cite{jensen-kato79}, where in the
radial case this term lives in the $\ell=1$ subspace (see \cite{jensen-kato79}
Remark~6.6).}
\end{remark}

\subsection{\it Resolvent expansions{\rm :} Proofs}

We now give some details on the proofs of the resolvent expansions.

\setcounter{theore}{9}
\begin{pot}
{\rm We give a brief outline of the proof. Since by assumption $M_0$ is invertible in
$\cK$, and since we assume $\beta>2p+3$, we can compute the inverse of $M(\ka)$ up to
an error term $\cO(\kappa^{p+1})$ by using the Neumann series and the expansion
\eqref{M-asymp}. This expansion is then inserted  into \eqref{freq}, leading to the
existence of the expansion up to terms of order $p$, and to the two expressions
\begin{equation*}
G_0=G_0^D-\goo v^*M_0^{-1}v\goo
\end{equation*}
and
\begin{equation*}
G_1=(I-\goo v^*M_0^{-1}v)G_1^D(I-v^*M_0^{-1}v\goo).
\end{equation*}
Now we carry out the following computation:
\begin{align*}
I-\goo v^*M_0^{-1}v&=I-\goo v^*(U+v\goo v^*)^{-1}v\\[.3pc]
&=I-\goo v^* U (I+v\goo v^*U)^{-1}v\\[.3pc]
&=I-\goo v^* U v(I+\goo v^*Uv)^{-1}\\[.3pc]
&=I-\goo V (I+\goo V)^{-1}\\[.3pc]
&=(I+\goo V)^{-1}.
\end{align*}
Using this result, and its adjoint, we get the expressions in the theorem. It is easy
to check that the above computations make sense between the weighted spaces.}
\end{pot}

\begin{pot}
{\rm We assume that zero is an exceptional point of the first kind. Thus we have that
$\dim\ker M_0=1$. Take $f\in\ker M_0$, $\|f\|=1$. Let $S=\ket{f}\bra{f}$ be the
orthogonal projection onto $\ker M_0$. Assume $\beta>2p+7$. Let $q=p+2$. Then by
Proposition~\ref{prop1} we have an expansion
\begin{equation}\label{ex1a}
M(\ka)=\sum_{j=0}^q\ka^jM_j+\cO(\ka^{q+1})=M_0+\ka \widetilde{M}_1(\ka).
\end{equation}
We now use Corollary 2.2 of \cite{JenNen}. Thus $M(\ka)$ is invertible, if and only if
\begin{equation}\label{ex1b}
m(\ka)=
\sum_{j=0}^{\infty}(-1)^j
\kappa^jS\left[\widetilde{M}_1(\kappa)J_0\right]^{j+1}S,
\end{equation}
is invertible as an operator on $S\cK$. We also recall the formula for the inverse
from\break Corollary~2.2 of \cite{JenNen},
\begin{equation}\label{M-1}
M(\ka)^{-1}=(M(\ka)+S)^{-1}+\frac{1}{\ka}
(M(\ka)+S)^{-1}Sm(\ka)^{-1}S(M(\ka)+S)^{-1}.
\end{equation}
It is easy to see that we have an expansion
\begin{equation*}
m(\ka)=\sum_{j=0}^{q-1}\ka^jm_j+\cO(\ka^q),
\end{equation*}
where
\begin{align}\label{am0}
m_0&=SM_1S,\\[.3pc]
m_1&=SM_2S-SM_1J_0M_1S,\label{am1}\\[.3pc]
m_2&=SM_3S-SM_1J_0M_2S-SM_2J_0M_1S
+SM_1J_0M_1J_0M_1S.\label{am2}
\end{align}
Using \eqref{dg1} we see that
\begin{equation}\label{SM1S}
m_0=SM_1S=-\ket{Svr}\bra{Svr}=-|\ip{f}{vr}|^2S.
\end{equation}
Since $\ip{f}{vr}\neq0$, it follows that $m_0$ is invertible in $S\cK$. The Neumann
series then yields an expansion
\begin{equation*}
m(\ka)^{-1}=m_0^{-1}+\sum_{j=1}^{q-1}\ka^jA_j+\cO(\ka^q).
\end{equation*}
The coefficients $A_j$ are in principle computable, although the expressions
rapidly get very complicated. This expansion is inserted into \eqref{M-1}. We
also use the Neumann series to expand
\begin{equation*}
(M(\ka)+S)^{-1}=J_0+\sum_{j=1}^q\ka^j\widetilde{M}_j+\cO(\ka^{q+1}).
\end{equation*}
This leads to an expansion
\begin{equation*}
M(\ka)^{-1}=\frac{1}{\ka}Sm_0^{-1}S+\sum_{j=0}^{q-2}\ka^jB_j+\cO(\ka^{q-1}),
\end{equation*}
where we also used that $SJ_0=J_0S=S$. We now use \eqref{freq} together with the
expansion above and the expansion of $R_0^D(-\ka^2)$ from Proposition~\ref{prop1} to
conclude that we have an expansion
\begin{equation*}
R(-\ka^2)=-\frac{1}{\ka}G_0^Dv^*Sm_0^{-1}SvG_0^D +
\sum_{j=0}^{q-2}\ka^{j}G_j+\cO(\ka^{q-1}).
\end{equation*}
This concludes the proof of the theorem.}
\end{pot}

\setcounter{theore}{12}
\begin{pot}
{\rm Assume that zero is an exceptional point of the third kind
for $H$. Thus $\dim\ker M_0\geq2$, and there exists an $f\in\ker
M_0$ with $\ip{vr}{f}\neq0$. We repeat the computations in the
proof of Theorem~\ref{thm-ex1}, although the assumptions are
different. As above, $S$ denotes the orthogonal projection onto
$\ker M_0$. Given $p\geq0$, assume $\beta>2p+11$, and let $q=p+4$.
Then $\beta>2q+3$, and for this $q$ we have the expansion
\eqref{ex1a}. We also have the expansion \eqref{ex1b} and the
expressions for the first three coefficients given in \eqref{am0},
\eqref{am1} and \eqref{am2}, respectively. We have
\begin{equation*}
m_0=SM_1S=-\ket{Svr}\bra{Svr},
\end{equation*}
which by our assumption is a rank 1 operator. The orthogonal projection onto
$\ker m_0$ is given by
\begin{equation*}
S_1=S+\frac{1}{\alpha}\ket{Svr}\bra{Svr},\quad\alpha=\|Svr\|^2_{\cK},
\end{equation*}
and by assumption $S_1\neq0$. Now we use the main idea in \cite{JenNen}, the repeated
application of Corollary~2.2. Applying it once more, we get
\begin{align}\label{m-1}
m(\ka)^{-1}&=(m(\ka)+S_1)^{-1}\notag\\[.3pc]
&\quad\,+\frac{1}{\ka} (m(\ka)+S_1)^{-1}S_1q(\ka)^{-1}S_1(m(\ka)+S_1)^{-1}
\end{align}
with
\begin{align}\label{aqj}
q(\ka)&=q_0+\ka q_1 + \cdots+\cO(\ka^{q-1})\notag\\[.3pc]
&= S_1m_1S_1+\ka [ S_1m_2S_1-S_1m_1(m_0+S_1)^{-1}m_1S_1]\notag\\[.3pc]
&\quad\,+\cdots+\cO(\ka^{q-1}).
\end{align}
Here the $\cdots$ are terms, whose coefficients can be computed explicitly. We must
have that $q_0$ is invertible in $S_1\cK$. Otherwise, we can iterate the procedure,
leading to a singularity in the expansion of $R(-\ka^2)$ of type $\ka^{-j}$ with
$j\geq3$, contradicting the self-adjointness of~$H$. Thus we have
\begin{equation}\label{q-1}
q(\ka)^{-1}=q_0^{-1}-\ka q_0^{-1}q_1q_0^{-1} +\cdots+\cO(\ka^{q-1}).
\end{equation}

It remains to perform the back-substitution, and to compute the coefficients.
The back-substitution leads to
\begin{equation*}
R(-\ka^2)=\frac{1}{\ka^2}G_{-2}+\frac{1}{\ka}G_{-1}+\cdots+\cO(\ka^{q-4}),
\end{equation*}
with expressions
\begin{align}\label{G-2p}
G_{-2}&=-G_0^DvS_1q_0^{-1}S_1vG_0^D,\\[.3pc]\label{G-1p}
G_{-1}&=G_0^DvS_1q_0^{-1}S_1m_2S_1q_0^{-1}S_1vG_0^D\notag\\[.3pc]
&\quad\,-G_0^Dv(S-S_1q_0^{-1}S_1m_1)(m_0+S_1)^{-1}(S-m_1S_1q_0^{-1}S_1)vG_0^D.
\end{align}
These expressions can be simplified. The computations are similar to the ones
in \cite{JN-FGR}, although there are some differences. Let $P_0$ denote
the projection onto the eigenspace for eigenvalue zero for $H$.

Let us start by reformulating the result in Lemma~\ref{NL-lemma}. Let
\begin{equation}
T=-G_0^Dv^*S_1\quad\text{and}\quad \Tt=UvP_0.
\end{equation}
The operator $T$ is {\it a priori} only bounded from $\cK$ to
$\cH^{-s}$ for $s>1/2$, but \lref{NL-lemma} shows that it is
actually bounded from $\cK$ to $\cH$, with $\Ran\,{T}=P_0\cH$. We
also have that $\Tt$ is bounded from $\cH$ to $\cK$, with
$\Ran\,\Tt=S_1\cK$. Now \lref{g2lemma} implies that
\begin{equation}\label{TT}
T\Tt=P_0\quad\text{and}\quad \Tt T=S_1.
\end{equation}
The adjoint $T^{\ast}$ is the closure of the operator $-S_1vG_0^D$. These observations
lead to the result
\begin{equation}
S_1q_0^{-1}S_1=-\Tt\Tt^{\ast}.
\end{equation}
Now insert into \eqref{G-2p} to get
\begin{equation*}
G_{-2}=T\Tt\Tt^{\ast}T^{\ast}=P_0.
\end{equation*}
Then we note that
\begin{equation}\label{G-1p1}
G_0^DvS_1q_0^{-1}S_1m_2S_1q_0^{-1}S_1vG_0^D=0.
\end{equation}
This result holds, since $S_1m_2S_1=S_1G_3^Dv^*S_1=0$, as can be seen from the kernel
\eqref{dg3} and the condition \eqref{nonlocalltwo}, which holds for all functions in
the range of $S_1$. As for  the last term in  \eqref{G-1p}, from  \eqref{SM1S} and
\eqref{aqj} it follows that
\begin{gather}\label{m0+S}
(m_0+S_1)^{-1}=S_1-\frac{1}{\alpha^4}\ket{Svr}\bra{Svr},\\[.3pc]
(S-S_1q_0^{-1}S_1m_1)S_1=0.
\end{gather}
Define
\begin{equation}\label{resonancefcn}
\Psi_c=\frac{1}{\|Svr\|^2}(G_0^Dv\ket{Svr}-P_0VG_2^Dv\ket{Svr}).
\end{equation}
Then a computation shows that we have
\begin{equation}
G_{-1}=\ket{\Psi_c}\bra{\Psi_c}.
\end{equation}
This concludes the proof of Theorem~\ref{thm-ex3}.}
\end{pot}

\setcounter{theore}{11}
\begin{pot}
{\rm We will not give the details of the proof of this theorem. It follows along the
lines of the previous proofs. More precisely, if as above $S$ is the orthogonal
projection onto $\ker M_0$, then (\ref{M-1})--(\ref{am2}) hold true with $m_0=0$, and
the argument leading to the invertibility of $q_0$, (see \eqref{aqj}), gives the fact
that $M_1$ is invertible. Then expanding in \eqref{M-1} and carrying the computation
far enough, one finds the expressions in (\ref{ex2-G-2})--(\ref{ex2-G1}) for the first
four coefficients explicitly, which are of interest in connection with the Fermi
golden rule results below.}
\end{pot}

\section{A non-local potential example}\label{nonlocal-sect}
\setcounter{equation}{0}

We will illustrate Theorem~\ref{thm-ex3} by giving an explicit example, using a rank 2
perturbation. The example is constructed such that $H$ has zero as an exceptional
point of the third kind.

Let us define two functions in $L^2([0,\infty))$ as follows:
\begin{align*}
\phi_1(r)&={\begin{cases}
0, & \text{for $0<r\leq 3$ }\\
1, & \text{for $3<r< 4$ }\\
0, & \text{for $4\leq r< \infty$ }
\end{cases}},\\[.3pc]
\phi_2(r)&={\begin{cases}
0, & \text{for $0<r\leq 1$ }\\
1, & \text{for $1<r< 2$ }\\
-\frac{3}{5}, & \text{for $2\leq r\leq 3$ }\\
0, & \text{for $3<r< \infty$ }
\end{cases}}.
\end{align*}
We have
\begin{equation}\label{eq91}
\int_0^{\infty}r\phi_1(r)\d r\neq0\quad\text{and}\quad \int_0^{\infty}r\phi_2(r)\d
r=0.
\end{equation}
As our potential we take
\begin{equation}
V=-\frac{3}{10}\ket{\phi_1}\bra{\phi_1}-\frac{75}{28}\ket{\phi_2}\bra{\phi_2}.
\end{equation}
For the factorization we take $\cK=\bC^2$, and define $v\in\bo{\cH}{\cK}$ by
\begin{equation}
v(f)=\begin{bmatrix}
\sqrt{\frac{3}{10}}\ip{\phi_1}{f}\\[.4pc]
\sqrt{\frac{75}{28}}\ip{\phi_2}{f}
\end{bmatrix}.
\end{equation}
We let $U=-I$, where $I$ is the identity operator on $\cK$. Then we have
$V=v^*Uv$. Next we compute $M_0$. Direct computation shows that
\begin{equation*}
vG_0^Dv^*=I.
\end{equation*}
The constants in $V$ were chosen to obtain this result. Thus $M_0=0$. Take
\begin{equation*}
f_1=\begin{bmatrix}
1\\[.2pc]
0
\end{bmatrix}
\quad\text{and}\quad
f_2=\begin{bmatrix}
0\\[.2pc]
1
\end{bmatrix}.
\end{equation*}
Then
\begin{equation*}
\ip{vr}{f_1}\neq0\quad\text{and}\quad\ip{vr}{f_2}=0,
\end{equation*}
due to \eqref{eq91}. Thus zero is an exceptional point of the third kind for
$H$ with this potential.
We can also find the resonance function and an eigenfunction explicitly. An
eigenfunction is given by $-G_0^Dv^*f_2$. Carrying out the computations,
one finds after normalization
\begin{equation}
\Psi_{0}(r)
=\sqrt{\frac{375}{98}}{\begin{cases}
-\frac{2}{5}r, & \text{for $0<r\leq 1$ }\\[.2pc]
\frac12 r^2-\frac75 r+\frac12, & \text{for $1<r< 2$ }\\[.2pc]
-\frac{3}{10}r^2+\frac95 r-\frac{27}{10}, & \text{for $2\leq r\leq 3$ }\\[.2pc]
0, &\text{for $3<r< \infty$}
\end{cases}}.
\end{equation}
\begin{fig*}[h]
\hskip 4pc{\epsfbox{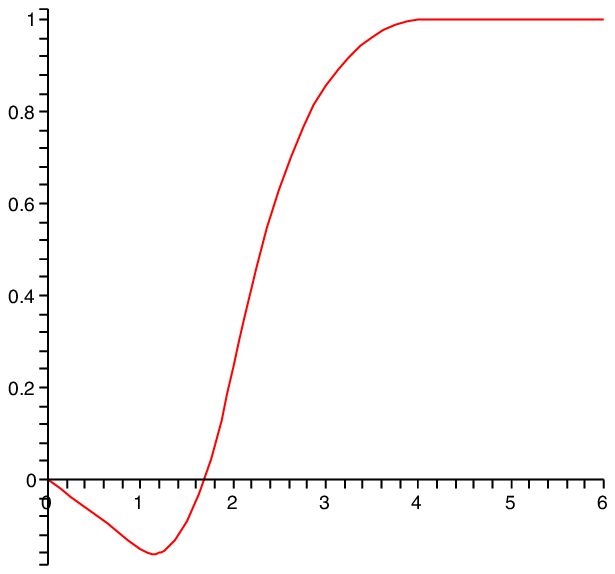}}\vspace{-.5pc} \caption{Canonical
zero resonance function $\Psi_c$.}\label{fig1}
\end{fig*}

\begin{fig*}
\hskip 4pc{\epsfbox{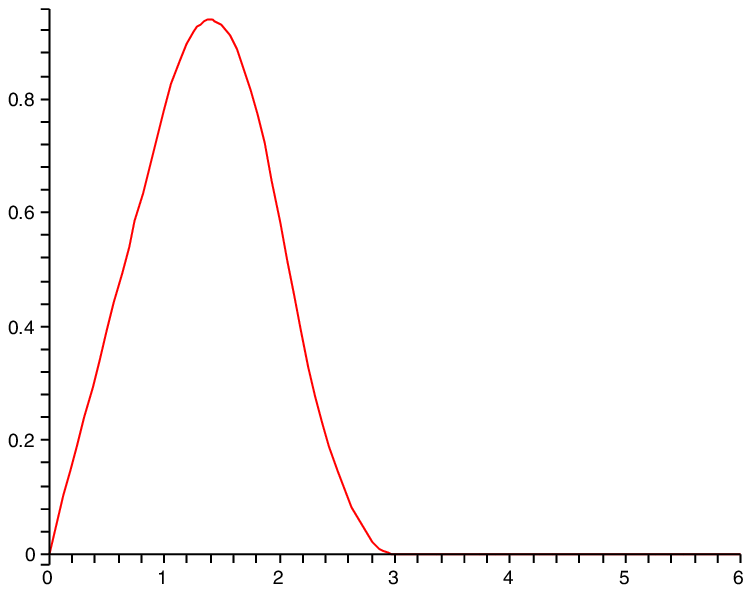}}\vspace{-.5pc} \caption{Normalized
zero eigenfunction $-\Psi_0$.}\label{fig2}\vspace{.5pc}
\end{fig*}

\noindent Using this function and the expression \eqref{resonancefcn} one gets
\begin{equation}
\Psi_c(r)=\begin{cases}
-\frac{52}{343}r, & \text{for $0<r\leq 1$ }\\[.4pc]
\frac{375}{686} r^2-\frac{61}{49} r+\frac{375}{686}, & \text{for $1<r< 2$ }\\[.4pc]
-\frac{225}{686}r^2+\frac{773}{343} r-\frac{2025}{686},
& \text{for $2\leq r\leq 3$ }\\[.4pc]
-\frac17 r^2 +\frac87 r -\frac97, & \text{for $3< r\leq 4$ }\\[.4pc]
1, & \text{for $4< r<\infty$ }
\end{cases}.
\end{equation}
The plots of the two functions are shown in figures~\ref{fig1} and
\ref{fig2}, respectively.

The computations in this example have been made using Maple, the computer algebra
system.

\section{Application to the Fermi golden rule at thresholds}\label{FGR}
\setcounter{equation}{0}

We recall the main result from \cite{JN-FGR} in the context of the
Schr\"{o}dinger operators on the half line considered above. Let $H=H_0^D+V$,
where $V$ satisfies Assumption~\ref{nonlocalV} for a sufficiently large $\beta$. Let $W$ be another potential
satisfying the same assumption. We consider the family $H(\ve)=H+\ve W$ for
$\ve>0$. Assume that $0$ is a simple eigenvalue of $H$, with normalized
eigenfunction $\Psi_0$. Assume
\begin{equation}\label{def-b}
b=\ip{\Psi_0}{W\Psi_0}>0.
\end{equation}
The results in \cite{JN-FGR} show that under some additional assumptions the
eigenvalue zero becomes a resonance for $H(\ve)$ for $\ve$ sufficiently small.
Here the concept of a resonance is the time-dependent one, as introduced in
\cite{orth}. The additional assumption needed is that for some odd integer
$\nu\geq-1$ we have
\begin{equation}\label{G-cond}
G_{j}=0,\quad\text{for $j=-1,1,\ldots,\nu-2$ \quad and}\quad
g_{\nu}=\ip{\Psi_0}{WG_{\nu}W\Psi_0}\neq0.
\end{equation}
Here $G_j$ denotes the coefficients in the asymptotic expansion for the
resolvent of $H$ around zero, as given in either Theorem~\ref{thm-ex2} or
Theorem~\ref{thm-ex3}. The main result in \cite{JN-FGR} gives the following
result on the survival probability for the state $\Psi_0$ under the evolution
$\exp(-itH(\ve))$. There exists $\ve_0>0$, such that for $0<\ve<\ve_0$
we have
\begin{equation}\label{FGRx1}
\ip{\Psi_0}{\e^{-itH(\ve)}\Psi_0}=\e^{-it\lambda(\ve)}+\delta(\ve,t),\quad t>0.
\end{equation}
Here $\lambda(\ve)=x_0(\ve)-i\Gamma(\ve)$ with
\begin{align}
\Gamma(\ve)&=-i^{\nu -1}g_{\nu}b^{\nu/2}\ve^{2+(\nu/2)}(1+\mathcal{O}(\ve)),\\[.3pc]
x_{0}(\ve)&=b\ve(1+\mathcal{O}(\ve)),
\end{align}
as $\ve\to0$. The error term satisfies
\begin{equation}\label{FGRx2}
|\delta(\ve,t)|\leq C \ve^{p(\nu)},\quad t>0, \quad p(\nu)=\min\{2,(2+\nu)/2\}.
\end{equation}
We state two corollaries to the results in this paper and in
\cite{JN-FGR}.

\begin{coro}$\left.\right.$\vspace{.5pc}

\noindent Let $H=H^D_0+V$ be a Schr\"{o}dinger operator on the
half line$,$ with $V$ satisfying  \aref{nonlocalV} for some
$\beta>17$. Assume that zero is an exceptional point of the second
kind for $H$. The zero eigenfunction is denoted by $\Psi_0$ and is
assumed to be simple. Let $W$ also satisfy \aref{nonlocalV} for
some $\beta>17$. Assume that
\begin{align}
b&=\ip{\Psi_0}{W\Psi_0}\neq0,\\[.3pc]
g_1&=\ip{\Psi_0}{WG_1W\Psi_0}\neq0.
\end{align}
Let $H(\ve)=H+\ve W$, $\ve>0$. The results \eqref{FGRx1}--\eqref{FGRx2} hold with
$\nu=1$.
\end{coro}
We note that an expression for $g_1$ can be obtained from \eqref{ex2-G1}.

\begin{coro}$\left.\right.$\vspace{.5pc}

\noindent Let $H=H^D_0+V$ be a Schr\"{o}dinger operator on the
half line{\rm ,} with $V$ satisfying  \aref{nonlocalV} for some
$\beta>9$. Assume that zero is an exceptional point of the third
kind for $H$. The zero eigenfunction is denoted by $\Psi_0$ and is
assumed to be simple. The canonical resonance function is denoted
by $\Psi_c$. Let $W$ also satisfy \aref{nonlocalV} for some
$\beta>9$. Assume that
\begin{align}
b&=\ip{\Psi_0}{W\Psi_0}\neq0,\label{cond-b}\\[.3pc]
g_{-1}&= \ip{\Psi_0}{WG_{-1}W\Psi_0}=|\ip{\Psi_0}{W\Psi_c}|^2\neq0.\label{cond-g-1}
\end{align}
Let $H(\ve)=H+\ve W$, $\ve>0$. The results \eqref{FGRx1}--\eqref{FGRx2} hold with
$\nu=-1$.
\end{coro}

This second Corollary is particularly interesting, since we can check the conditions
\eqref{cond-b} and \eqref{cond-g-1} in the example given in \S~\ref{nonlocal-sect}. It
is easy to see that one can get both $\ip{\Psi_0}{W\Psi_c}\neq0$ and
$\ip{\Psi_0}{W\Psi_c}=0$, for both local and non-local perturbations $W$. Only in the
first case can one apply directly the results from \cite{JN-FGR}, due to the condition
\eqref{G-cond}. The other case has not yet been investigated in detail.

One can also use the results on resolvent expansions to give examples using two
channel models, as in \cite{JN-FGR}. We omit stating these results explicitly.

\section*{Acknowledgments}

The first author (AJ) was partially supported by a grant from the Danish Natural
Sciences Research Council. The second author (GN) was partially supported by Aalborg
University and was also supported in part by CNCSIS under Grant 905-13A/2005.


\begin{thebibliography}{10}

\bibitem{aktosun00}
Aktosun T, Factorization and small-energy asymptotics for the radial Schr\"{o}dinger
equation, {\it J. Math. Phys.} {\bf 41} (2000) 4262--4270

\bibitem{amrein}
Amrein W O, Berthier A-M and Georgescu V, Lower bounds for zero
energy eigenfunctions of Schr\"{o}dinger operators, {\it Helv.
Phys. Acta} {\bf 57} (1984) 301--306

\bibitem{jensen80}
Jensen A, Spectral properties of {S}chr\"{o}dinger operators and
time-decay of the wave functions, {R}esults in
${L}^2(\mathbf{{R}}^m)$, $m\geq5$, {\it Duke Math. J.} {\bf 47}
(1980) 57--80

\bibitem{jensen-kato79}
Jensen A and Kato T, Spectral properties of {S}chr\"{o}dinger operators and time-decay
of the wave functions, {\it Duke Math. J.} {\bf 46} (1979) 583--611

\bibitem{JenNen}
Jensen A and Nenciu G, A unified approach to resolvent expansions at thresholds, {\it
Rev. Math. Phys.} {\bf 13} (2001) 717--754

\bibitem{JenNen-err}
Jensen A and Nenciu G, Erratum to the paper: A unified approach to resolvent
expansions at thresholds, {\it Rev. Math. Phys.} {\bf 16} (2004) 675--677

\bibitem{JN-FGR}
Jensen A and Nenciu G, The Fermi golden rule and its form at
thresholds in odd dimensions, {\it Comm. Math. Phys.} {\bf 261(3)}
(2006) 693--727

\bibitem{murata82}
Murata M, Asymptotic expansions in time for solutions of
{S}chr\"odinger-type equations, {\it J. Funct. Anal.} {\bf 49(1)}
(1982) 10--56

\bibitem{orth}
Orth A, Quantum mechanical resonance and limiting absorption: the
many body problem, {\it Comm. Math. Phys.} {\bf 126(3)} (1990)
559--573

\bibitem{rs1}
Reed M and Simon B, Methods of Modern Mathematical Physics I:
Functional Analysis, revised and enlarged edition (New York:
Academic Press) (1980)

\bibitem{yafaev82}
Yafaev D R, The low energy scattering for slowly decreasing potentials, {\it Comm.
Math. Phys.} {\bf 85} (1982) 177--196
\end{thebibliography}
\end{document}